\documentclass[preprint2]{aastex62}

\hypersetup{linkcolor=black,citecolor=black,filecolor=cyan,urlcolor=magenta}

\usepackage{amsmath}

\submitjournal{ApJL}

\shorttitle{SOFIA/FIFI-LS [C\,{\sc ii}] map of M51}
\shortauthors{Pineda, J.L. et al.}
\usepackage{epstopdf}

\begin{document}
\title{A SOFIA Survey of [C\,\sc ii] in the galaxy M51 I. [C\,{\sc ii}]
  as a tracer of Star Formation}

\author[0000-0001-8898-2800]{Jorge L. Pineda}
\affiliation{Jet Propulsion Laboratory, California Institute of Technology, 4800 Oak Grove Drive, Pasadena, CA 91109-8099, USA} 
\author[0000-0003-2649-3707]{Christian Fischer}
\affiliation{Deutsches SOFIA Institut, Pfaffenwaldring 29, 70569 Stuttgart, Germany}
\author[0000-0001-7647-7348]{Maria Kapala}
\affiliation{University of Cape Town, Department of Astronomy, Private Bag X3, 7701 Cape Town, Republic Of South Africa}
\author{J\"urgen Stutzki}
\affiliation{KOSMA, I. Physikalisches Institut, Universit\"at zu K\"oln, Z\"ulpicher Stra\ss e 77, 50937\\ K\"oln, Germany} 
\author{Christof Buchbender}
\affiliation{KOSMA, I. Physikalisches Institut, Universit\"at zu K\"oln, Z\"ulpicher Stra\ss e 77, 50937\\ K\"oln, Germany} 
\author[0000-0002-6622-8396]{Paul F. Goldsmith}
\affiliation{Jet Propulsion Laboratory, California Institute of Technology, 4800 Oak Grove Drive,\\ Pasadena, CA 91109-8099, USA} 
\author{Monika Ziebart}
\affiliation{KOSMA, I. Physikalisches Institut, Universit\"at zu K\"oln, Z\"ulpicher Stra\ss e 77, 50937\\ K\"oln, Germany} 
\author{Simon C. O. Glover}
\affiliation{Universit\"at Heidelberg, Zentrum f\"ur Astronomie, Albert-Ueberle-Str. 2, D-69120 Heidelberg, Germany}
\author[0000-0002-0560-3172]{Ralf S. Klessen}
\affiliation{Universit\"at Heidelberg, Zentrum f\"ur Astronomie, Albert-Ueberle-Str. 2, D-69120 Heidelberg, Germany and Universit\"at Heidelberg, Interdisziplin\"ares Zentrum f\"ur Wissenschaftliches Rechnen, INF 205, D-69120 Heidelberg, Germany}
\author[0000-0002-8762-7863]{Jin Koda}
\affiliation{Department of Physics and Astronomy, Stony Brook University, Stony Brook, NY 11794-3800, USA} 
\author{Carsten Kramer}
\affiliation{Instituto Radioastronomía Milim\'etrica (IRAM), Av. Divina Pastora 7, Nucleo Central, 18012, Granada, Spain}
\author{Bhaswati Mookerjea}
\affiliation{Tata Institute of Fundamental Research, Homi Bhabha Road, Mumbai, 400005, India}
\author[0000-0002-4378-8534]{Karin Sandstrom}
\affiliation{Center for Astrophysics and Space Sciences, University of California,\\ San Diego, CA, USA}
\author[0000-0002-0438-3323]{Nick Scoville}
\affiliation{California Institute of Technology, MC 249-17, 1200 East California Boulevard, Pasadena, CA 91125, USA }
\author[0000-0002-0820-1814]{Rowan Smith}
\affiliation{Jodrell Bank Centre for Astrophysics, School of Physics and Astronomy, University of Manchester, Oxford Road, Manchester M13 9PL, UK}

\correspondingauthor{Jorge L. Pineda}
\email{Jorge.Pineda@jpl.nasa.gov}

\begin{abstract}

  We present a [C\,{\sc ii}] 158\,$\mu$m map of the entire M51
  (including M51b) grand--design spiral galaxy observed with the
  FIFI--LS instrument on SOFIA. We compare the [C\,{\sc ii}] emission
  with the total far--infrared (TIR) intensity and star formation rate
  (SFR) surface density maps (derived using H$\alpha$ and 24$\mu$m
  emission) to study the relationship between [C\,{\sc ii}] and the
  star formation activity in a variety of environments within M51 on
  scales of 16\arcsec\ corresponding to $\sim$660\,pc. We find that
  [C\,{\sc ii}] and the SFR surface density are well correlated in the
  central, spiral arm, and inter-arm regions. The correlation is in
  good agreement with that found for a larger sample of nearby
  galaxies at kpc scales. We find that the SFR, and [C\,{\sc ii}] and
  TIR luminosities in M51 are dominated by the extended emission in
  M51's disk.  The companion galaxy M51b, however, shows a deficit of
  [C\,{\sc ii}] emission compared with the TIR emission and SFR
  surface density, with [C\,{\sc ii}] emission detected only in the
  S--W part of this galaxy.  The [C\,{\sc ii}] deficit is associated
  with an enhanced dust temperature in this galaxy. We interpret the
  faint [C\,{\sc ii}] emission in M51b to be a result of suppressed
  star formation in this galaxy, while the bright mid-- and
  far--infrared emission, which drive the TIR and SFR values, are
  powered by other mechanisms. A similar but less pronounced effect is
  seen at the location of the black hole in M51's center.  The
  observed [C\,{\sc ii}] deficit in M51b suggests that this galaxy is
  a valuable laboratory to study the origin of the apparent [C\,{\sc
    ii}] deficit observed in ultra--luminous galaxies.
\end{abstract}



\section{Introduction} \label{sec:intro}

The [C\,{\sc ii}] 158\,$\mu$m line is the main coolant of diffuse
neutral interstellar gas and therefore plays a critical role in the
thermal balance of the interstellar medium of galaxies
\citep{Dalgarno1972}.  Interstellar gas is heated by energetic
electrons that are emitted by dust grains that absorb far--ultraviolet
photons from massive stars \citep{Spitzer1948}.  [C\,{\sc ii}]
emission is thus related to the energy input from massive stars to the
interstellar medium, making it the brightest far-infrared line
\citep{Stacey1991} and an important tracer of the star formation
activity in galaxies. Redshifted [C\,{\sc ii}] line emission has been
used to study the properties of unresolved galaxies in the early
Universe with ALMA and NOEMA \citep{Carilli2013}. It is however
important to study the relationship between star formation and the
[C\,{\sc ii}] emission in the local Universe, where different
environments (nuclei, spiral arms, etc.) can be separated, and their
relative contribution to the observed star formation tracers can be
determined.

The [C\,{\sc ii}] emission and star--formation rate have been observed
to be well correlated in the Milky Way \citep{Pineda2014}, within
nearby galaxies \citep{Herrera-Camus2015}, in the integrated emission
of unresolved nearby galaxies \citep{deLooze2011}, and even in a
sample of ``normal'' high-redshift galaxies
\citep{Capak2015}. However, large [C\,{\sc ii}] intensity deficits
(reduced [C\,{\sc ii}] emission for a given far-infrared (FIR)
intensity) have been observed in ultra-luminous infrared galaxies
\citep[e.g.][]{Malhotra2001,Diaz-Santos2013,Diaz-Santos2017}, which
has been interpreted as [C\,{\sc ii}] failing to trace the (possibly
enhanced) star formation activity in these galaxies. The origin of
this deficit remains elusive, with several possibilities being
suggested in the literature that might reduce the [C\,{\sc ii}]
intensity and/or increase the FIR intensity
\citep[e.g.][]{Gracia-Carpio2011,Kapala2017,Diaz-Santos2013,
  Goicoechea2015,Langer2015,Smith2017}. Observations of nearby
galaxies provide an important opportunity to study the origin of the
[C\,{\sc ii}] deficit and its relationship to the different
environments in these galaxies in detail.

In this Letter we present a complete [C\,{\sc ii}] map of the M51
galaxy observed with the FIFI-LS instrument on SOFIA. This map was
obtained as part of a Joint Impact Proposal (program ID {\tt
  04\_0116}) that also includes a velocity resolved map of M51
obtained with the upGREAT instrument on SOFIA. In this Letter, we use
the FIFI-LS [C\,{\sc ii}] map to study the relationship of this
spectral line to star formation over the entire disk of the M51,
including its companion M51b (NGC5195). The velocity--resolved
spectral map of M51 will be presented in a separate paper with
emphasis on the gas kinematics.

M51 is a nearby grand design spiral at a distance of 8.5\,Mpc
\citep{McQuinn2016}, with an inclination angle of 24\degr\
\citep{Daigle2006}. It is interacting with a smaller companion galaxy,
M51b , classified as a barred lenticular (SB0 pec;
\citealt{Sandage1981}) and a LINER galaxy \citep{Ho1997}.  Partial
maps and individual positions of M51 have been presented based on
observations with the KAO \citep{Nikola2001} and ISO
\citep{Kramer2005} at resolutions of 55\arcsec\ and 80\arcsec,
respectively, which were insufficient to separate different
environments within the galaxy. \citet{Parkin2013} presented [C\,{\sc
  ii}] observations at 12\arcsec\ angular resolution in M51 using the
{\it Herschel}/PACS instrument, but focused only on the inner parts of
M51.  With the complete [C\,{\sc ii}] map of the M51 and M51b galaxies
we are able to study the relationship between the [C\,{\sc ii}]
emission and the star formation activity in a wide range of
environments.

This paper is organized as follows. In Section~\ref{sec:obs} we detail
the FIFI-LS observations and data reduction. We also describe the
ancillary data we use to determine the star--formation rate (SFR) and
total far-infrared intensity (TIR) in M51. In
Section~\ref{sec:results} we compare the [C\,{\sc ii}], TIR
intensities, and the SFR surface density in M51. We discuss the
results of this comparison in Section~\ref{sec:discussion}, and give
the conclusions of this work in Section~\ref{sec:conclusions}.

\section{Observations} \label{sec:obs}

\begin{figure}[t] 
\includegraphics[width=0.53\textwidth]{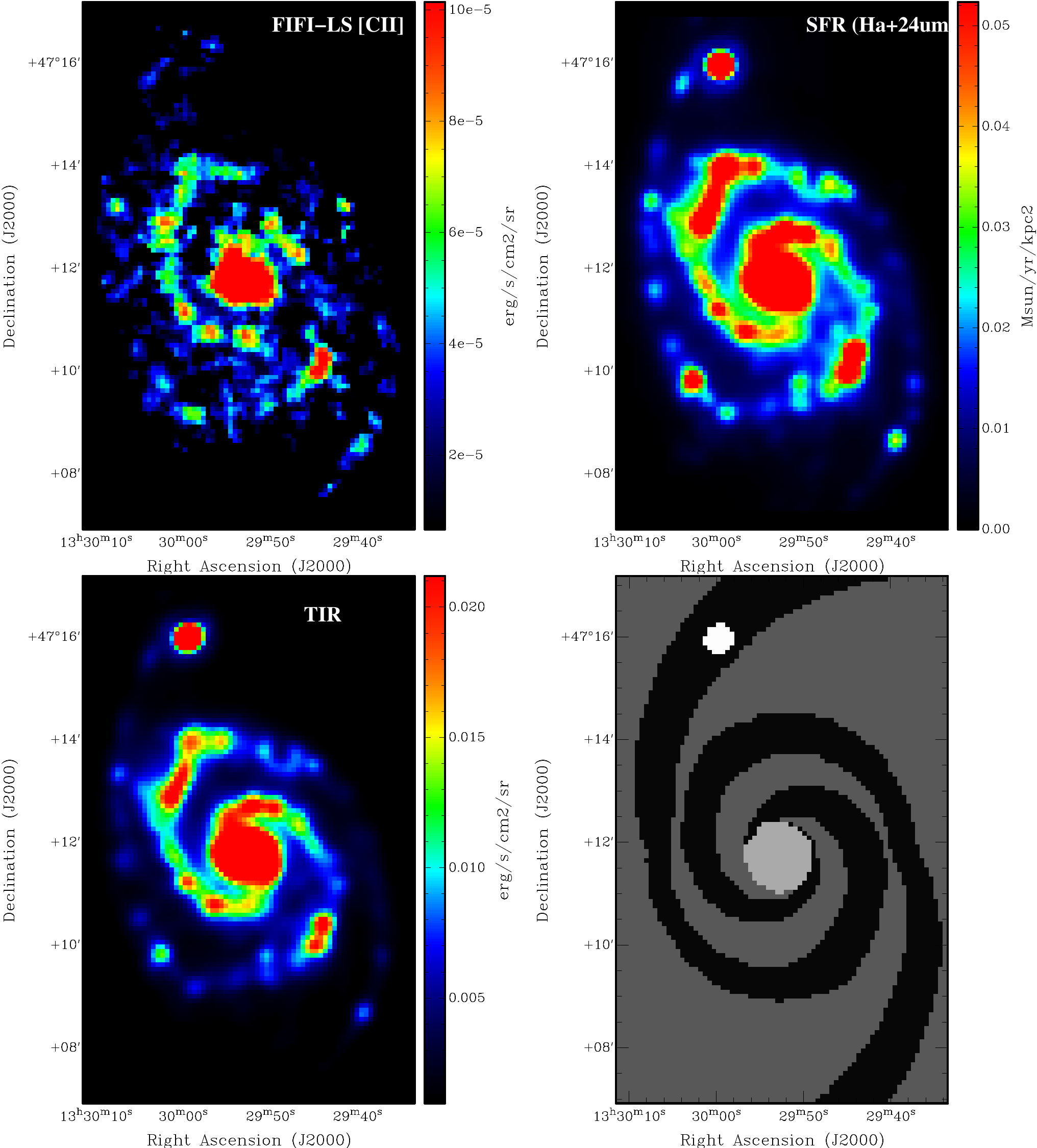}
\caption{Images of the galaxy M51 in [C\,{\sc ii}], the star formation
  rate surface density ($\Sigma_{\rm SFR}$), and the total infrared
  intensity (TIR). We also include a mask image showing the different
  environments studied here, center, arms, inter-arms, and M51b,
  indicated by light--grey, black, dark--grey, and white regions,
  respectively. \label{fig:m51_maps} }
\end{figure}
\subsection{FIFI-LS [C\,{\sc ii}]  observations} 

We present data taken on the [C\,{\sc ii}] line around 157.8\,$\mu$m,
with the Far Infrared Field-Imaging Line Spectrometer (FIFI--LS)
instrument \citep{Colditz2012,Klein2014a} aboard the Stratospheric
Observatory For Infrared Astronomy \citep[SOFIA;][]{Young2012}.
Details on the FIFI--LS instrument, observing schemes, and atmospheric
correction can found in \citet{Fischer2018}.  Here we present data
taken with the 115--203\,$\mu$m channel that has a 5$\times$5 pixel
footprint on the sky, each with a size of 12\arcsec\ $\times$
12\arcsec\ yielding a field--of--view of 1\arcmin $\times
$1\arcmin. Each pixel is a so-called ``spaxel'', which means that
internally the light is dispersed spectrally (using a grating) over 16
pixels for each spaxel, providing an integral-field data cube covering
a total spectral bandwidth between 1500 and 3000\,km\,s$^{-1}$, with a
spectral resolution of 270\,km\,s$^{-1}$.  Our data were acquired in
symmetric chop mode with a full throw of 8\arcmin. The M51 map is a
mosaic of 181 fields of 1\arcmin $\times$ 1\arcmin\ in a grid with
30\arcsec\ spacing to create half--pixel sampling and some redundancy
in the data set. The on--source integration time per point is
120s. With the overheads for chopping and telescope motion, the whole
map took about 15\,h. We assume a diffraction--limited telescope with
a point spread function (PSF) having a FWHM of 16\arcsec\ (660\,pc at
a distance of 8.5\,Mpc). The data were taken on 11 SOFIA flights in
two series between February 2016 and March 2017. Data reduction was
carried out using the FIFI-LS data reduction pipeline
\citep{Vacca2016}.  We fitted a linear baseline to each spectrum
  and integrated the emission within the [C\,{\sc ii}] spectral region
  to obtain the integrated intensity map. final [C\,{\sc ii}]
map has a pixel size of 5.3\arcsec\ and has a typical r.m.s. noise of
$6.6\times10^{-6}$\,erg\,s$^{-1}$\,cm$^{-2}$\,sr$^{-1}$.

Flux calibration is established by observing standard calibration
sources.  The absolute amplitude calibration uncertainty is assumed to
be 20\%, of which 10\% is the relative uncertainty from FIFI-LS seen
between flight series and different calibrators, and the rest is the
uncertainty in the atmospheric transmission correction.  We compared
the intensities of our FIFI-LS map in the inner parts of M51 to that
observed with {\it Herschel}/PACS \citep{Parkin2013} smoothed to an
angular resolution that matched that of FIFI--LS.  We found that the
FIFI-LS fluxes are systematically lower by a factor of $\sim$1.26
compared with those from PACS.  While this factor is within the
uncertainties of our observations, we nevertheless apply this
correction factor to the FIFI--LS intensities used in the analysis
presented here.

\subsection{TIR and SFR maps} 

We used {\it Spitzer} and {\it Herschel} mid and far--infrared
continuum maps \citep{Kennicutt2011} to derive the TIR intensity
($I_{\rm TIR}$) and SFR surface density ($\Sigma_{\rm SFR}$) in
M51. We estimated the TIR intensity (i.e. the infrared intensity
integrated between 3 and 1100\,$\mu$m; in units of erg s$^{-1}$
cm$^{-2}$\,sr$^{-1}$) using,

\begin{multline}
  I_{\rm TIR} = 
0.95 \nu I_{\nu, \rm 8\mu m} + 1.15 \nu I_{\nu, \rm 24\mu m} \\+ \nu I_{\nu, \rm 70\mu m} + \nu I_{\nu, \rm 160\mu m},
\end{multline}

where all specific intensities, $I_\nu$, are in units of in
erg\,s$^{-1}$\,cm$^{-2}$\,sr$^{-1}$\,Hz$^{-1}$ \citep{Croxall2012}. The
surface density of recent star formation (in units of
M$_{\odot}$\,yr$^{-1}$\,kpc$^{-2}$) was estimated using 24$\mu$m and
H$\alpha$ maps following \citet{Gallagher2018},
\begin{equation}
\Sigma_{SFR} =  634 I_{\rm H\alpha} + 0.0025 I_{24\mu m},
\end{equation} 
where the intensity of $I_{\rm H\alpha}$ is in units of
erg\,s$^{-1}$\,cm$^{-2}$\,sr$^{-1}$, and $I_{24 \mu m}$ in units of
MJy\,sr$^{-1}$. The H$\alpha$ and 24$\mu$m continuum maps were
observed as part of the {\it Spitzer}/SAGE survey
\citep{Kennicutt2003}.  Both maps were smoothed with a Gaussian kernel
and regridded to match the resolution and grid of the FIFI-LS [C\,{\sc
  ii}] map.  We estimated the rms noise of $I_{\rm TIR}$ and
  $\Sigma_{\rm SFR}$ by calculating the standard deviation in regions
  in the maps that are removed from the galaxy. The rms noise is
  $1.3\times10^{-3}$\,M$_{\odot}$\,yr$^{-1}$\,kpc$^{-2}$ for the SFR
  surface density, and $2.5 \times 10^{-4}$ erg s$^{-1}$
  cm$^{-2}$\,sr$^{-1}$ for the TIR intensity.

\begin{figure*}[t]
\includegraphics[width=0.75\textwidth]{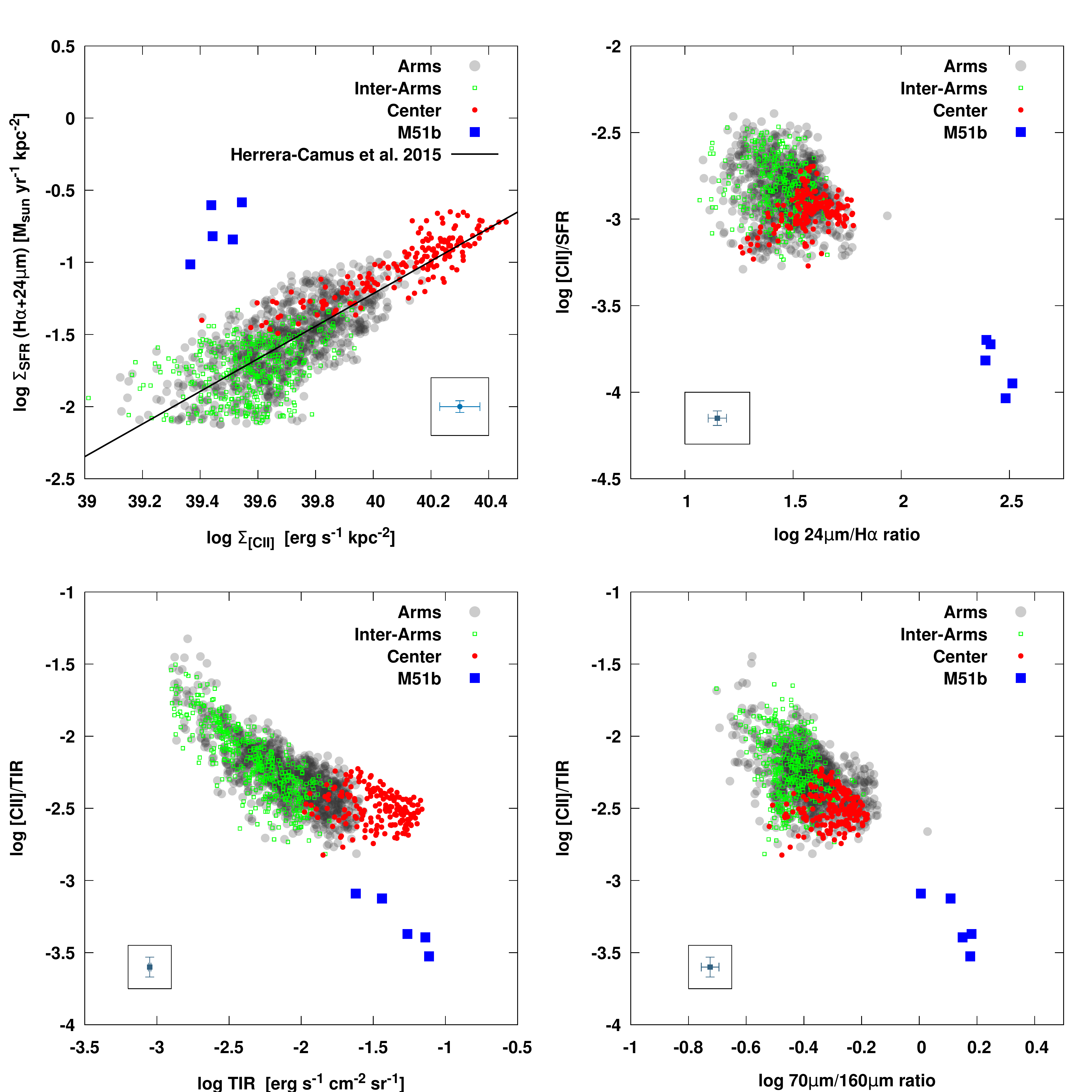}
\caption{({\it Upper left panel}) The star formation rate surface
  density as a function of the [C\,{\sc ii}] luminosity per unit
  area. The straight line corresponds to the fit of this relationship
  obtained for a sample of 46 nearby galaxies presented by
  \citet{Herrera-Camus2015}. ({\it Lower left panel}) Ratio of
  [C\,{\sc ii}] to TIR intensity as a function of the TIR intensity.
  ({\it Upper right panel}) The ratio of the [C\,{\sc ii}] luminosity
  per unit area to the star formation rate surface density as a
  function of the 24$\mu$m to H$\alpha$ ratio.  ({\it Lower right
    panel}) Ratio of [C\,{\sc ii}] to TIR intensity as a function of
  the 70$\mu$m/160$\mu$m ratio.  In all plots different environments
  defined in Figure~\ref{fig:m51_maps} are color coded. Typical error
  bars are shown inside a box.  All data points shown are above the
  5$\sigma$ level in each quantity. \label{fig:cii_vs_tir_sfr} }
\end{figure*}

\begin{deluxetable}{lcccc}[t]
\tablecaption{M51 Luminosities \label{tab:obs}}
\tablecolumns{5}
\tablewidth{0pt}
\tablehead{
\colhead{Region} &
\colhead{\# of} & \colhead{$L$([C\,{\sc ii}]) } & \colhead{$L$(TIR)} & \colhead{SFR}\\
\colhead{} & \colhead{ pixels\tablenotemark{a} } &
\colhead{[erg/s]} & \colhead{[erg/s]} & \colhead{[$M_{\odot}$/yr]}
}
\startdata
Center     & 182   & 1.2$\times 10^{41}$   & 3.8$\times 10^{43}$ & 0.9  \\ 
Arms       & 2865  & 2.5$\times 10^{41}$   & 9.0$\times 10^{43}$ & 2.5  \\ 
Inter--arm & 7382  & 8.5$\times 10^{40}$   & 4.7$\times 10^{43}$ & 1.2  \\ 
M51b      & 35    & 7.0$\times 10^{38}$   & 8.4$\times 10^{42}$ & 0.2  \\ 
\hline                                   
Total      & 10464 & 4.5$\times 10^{41}$   & 1.8$\times 10^{44}$ & 4.8  \\ 
\enddata
\tablenotetext{a}{Total number of pixels in mask.}
\end{deluxetable}

\section{Results} \label{sec:results}

In Figure~\ref{fig:m51_maps} we show the FIFI-LS [C\,{\sc ii}] map of
M51 together with those of $\Sigma_{\rm SFR}$ and $I_{\rm TIR}$.  We
also include a mask that is used to separate different environments in
M51.  These are the center, arms, inter-arms, and the M51b companion
galaxy. The arm regions were defined following a geometrical model of
the spiral structure in M51 (Pineda et al. 2018, in prep.) and the
center and M51b masks were defined as pixels with a K--band magnitude
(tracing the stellar mass) larger than 17.2\,mag and 16\,mag,
respectively, at 16\arcsec\ resolution.  All three images are
morphologically similar, showing peaks at the center, northern, and
southern spiral arms. The only exception is the M51b galaxy that is
bright in the $\Sigma_{\rm SFR}$ and $I_{\rm TIR}$ maps but much
fainter in the [C\,{\sc ii}] map.


The upper left panel of Figure~\ref{fig:cii_vs_tir_sfr} shows a
pixel-by-pixel comparison between the [C\,{\sc ii}] luminosity per
unit area observed in M51 and the SFR surface density. Data points
from the different mask regions defined in Figure~\ref{fig:m51_maps}
(arms, inter--arms, center, and M51b) are color coded.  We also show a
straight line that corresponds to the $\Sigma_{\rm SFR}-\Sigma_{\rm
  [CII]}$ relationship found by \citet{Herrera-Camus2015} in a sample
of 46 nearby galaxies observed with {\it Herschel}/PACS. While the
[C\,{\sc ii}] intensity range of the center, spiral arms, and
inter-arm regions in M51 differ, they all show a good correlation
between $\Sigma_{\rm SFR}$ and $\Sigma_{\rm [CII]}$.  The data points
in these regions are also in agreement with the $\Sigma_{\rm SFR} -
\Sigma_{\rm [CII]}$ relationship of the galaxy sample studied by
\citet{Herrera-Camus2015}.  However, M51b shows significantly fainter
[C\,{\sc ii}] emission with respect to the SFR surface density,
suggesting a [C\,{\sc ii}] deficit similar to that seen in
ultra-luminous infrared galaxies (ULIRGS). A moderate deficit is also
seen in pixels at M51's center.

The lower left panel of Figure~\ref{fig:cii_vs_tir_sfr} shows the
[C\,{\sc ii}] to TIR intensity ratio as a function of the TIR
intensity. The data points center, arm, and inter--arm regions are
typically in the 10$^{-3}$ to 10$^{-2}$ range observed in normal
galaxies \citep[e.g.][]{Kapala2015,Kramer2013}.  M51b, exhibits much
lower values of the [C\,{\sc ii}]/TIR, as low as $\sim$10$^{-4}$. Such
low values are typical of ULIRGS \citep{Diaz-Santos2013}, but note
that the TIR luminosity of M51b (Table \ref{tab:obs}) is at least two
orders of magnitude lower than that of typical ULIRGS.

The upper right panel of Figure~\ref{fig:cii_vs_tir_sfr} compares the
$\Sigma_{\rm [CII]}/\Sigma_{\rm SFR}$ ratio to the 24$\mu$m/H$\alpha$
ratio to study whether there are systematic differences between star
forming regions in the arms and more diffuse regions in the inter-arms
of M51. We find no significant difference between the
24$\mu$m/H$\alpha$ in the arm (average ratio 31.5) and inter-arm
(average ratio 27.5) regions of M51.  The central region has a
somewhat larger 24$\mu$m/H$\alpha$ ratio of 39.  However, M51b
deviates significantly with an average 24$\mu$m/H$\alpha$ ratio of
276. This deviation is a result of enhanced 24$\mu$m emission rather
than fainter H$\alpha$ emission.   We show in the lower right
  panel of Figure~\ref{fig:cii_vs_tir_sfr}, the [C\,{\sc ii}]/TIR
  ratio as a function of the 70$\mu$m/160$\mu$m ratio, which is a
  proxy for the dust temperature. The arm and inter-arm regions have
  similar average 70$\mu$m/160$\mu$m ratio (0.35 and 0.4,
  respectively) while the central region has a slightly larger average
  value (0.5). M51b shows an enhanced average 70$\mu$m/160$\mu$m ratio
  of 1.3, which indicates that the reduced [C\,{\sc ii}]/TIR ratio in
  this galaxy is related to an increase in dust temperature.  The
  suggested enhancement of the dust temperature in M51b is consistent
  with that estimated by \citet{MentuchCooper2012}.  A similar
  dependence of the [C\,{\sc ii}]/TIR ratio to dust temperature seen
  in Figure~\ref{fig:cii_vs_tir_sfr} has been observed in a larger
  sample of galaxies \citep[including ULIRGS;
  ][]{Herrera-Camus2015,Diaz-Santos2017,Diaz-Santos2013,Lu2015}.

\begin{figure}[t]
\plotone{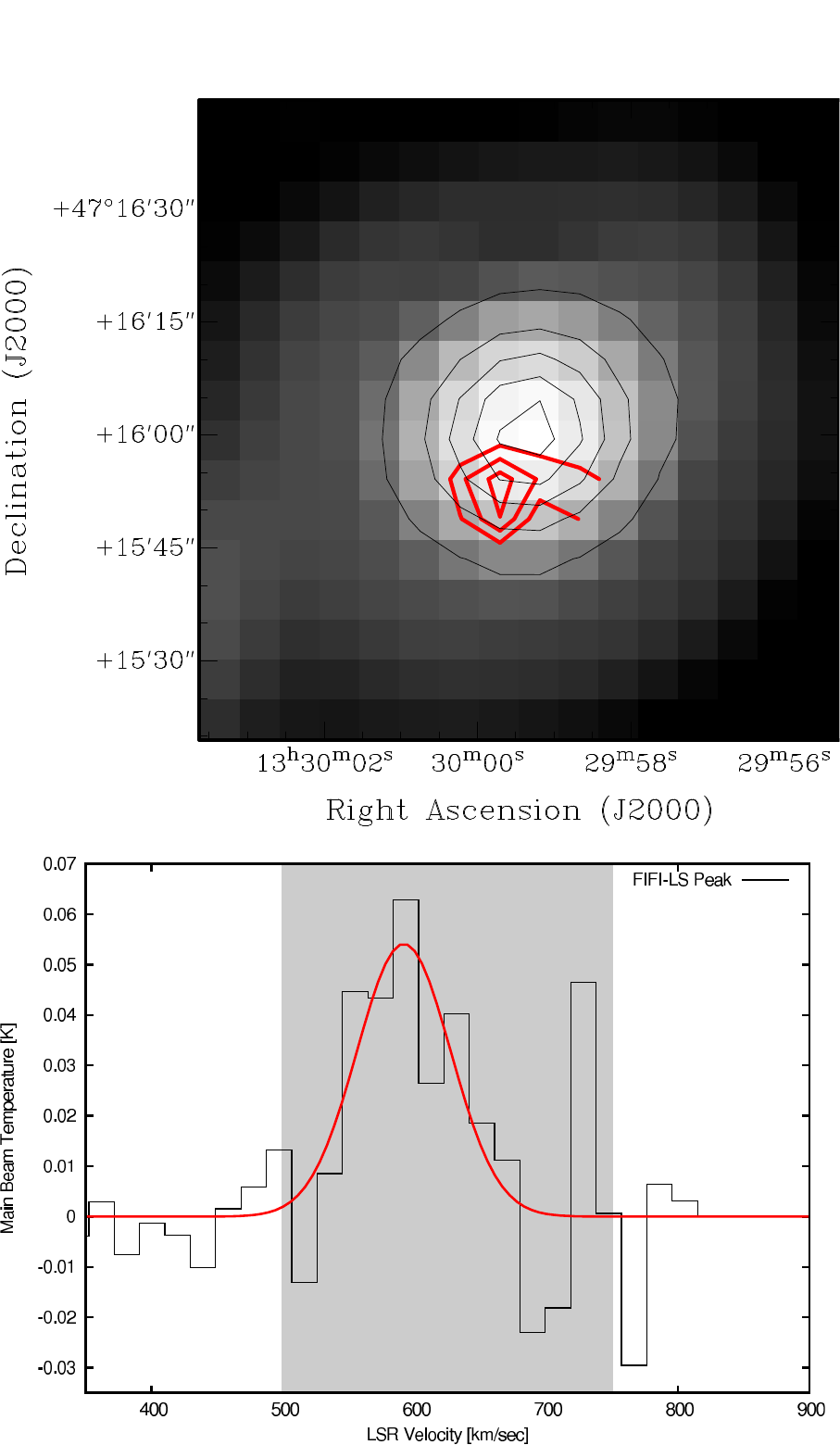}
\caption{({\it Upper panel}) Image of M51b observed in [C\,{\sc ii}]
  with SOFIA/FIFI-LS. The red contours correspond to 65\% of the peak
  (2.9$\times10^{-5}$ erg\,s$^{-1}$\,cm$^{-2}$\,sr$^{-1}$) to 100\% in
  steps of 15\%.  The grey-scale and black contours indicate the TIR
  intensity. ({\it Lower panel}) [C\,{\sc ii}] spectrum observed with
  the SOFIA/upGREAT instrument resulting from the average of the
  spectra above the 2.0$\times10^{-5}$
  erg\,s$^{-1}$\,cm$^{-2}$\,sr$^{-1}$ 3$\sigma$ level in the FIFI-LS
  image and S--W from M51b's center. The red line denotes the result
  of a Gaussian fit and the grey area the expected velocity range of
  the integrated CO emission observed in M51b
  \citep{Kohno2002a}. \label{fig:m51b_zoom} }
\end{figure}

\section{Discussion}\label{sec:discussion}

\subsection{Relative contributions from M51 environments}
\label{sec:relat-contr-m51}

Table\,\ref{tab:obs} presents the total [C\,{\sc ii}] and TIR
luminosities, and SFR for the different environments studied in
M51. The SFR and the [C\,{\sc ii}] and TIR luminosities in M51 are
dominated by the disk of M51 (arm and inter-arm regions) representing
about 75\% of the total value of these quantities, even though the
central region has the highest surface density. The larger luminosity
in the disk of M51 is a result of these quantities being extended over
a larger area than that of the central and M51b regions. This suggests
that if M51 were at a greater distance, and thus were unresolved, the
measured SFR and [C\,{\sc ii}] and TIR luminosities would be dominated
by the galaxy's disk rather than by its central region.

\subsection{The origin of the [C\,{\sc ii}] deficit in M51b}
\label{sec:origin-c-sc}

In the upper panel of Figure~\ref{fig:m51b_zoom} we show [C\,{\sc ii}]
emission from M51b in the form of contours. We also show the TIR
emission.  While most of the TIR emission (and inferred SFR) peaks at
the center of the galaxy, [C\,{\sc ii}] shows emission only from the
South--West (S--W) side of the TIR peak. CO observations show a
molecular disk in M51b that extends to the S--W and North--East (N--E)
from its center \citep{Kohno2002a,Alatalo2016}. The S--W and N--E
components have CO velocities of $\sim$580 km\,s$^{-1}$ and $\sim$680
km\,s$^{-1}$, respectively \citep{Kohno2002a}. We also observed this
region with the upGREAT instrument with velocity resolved observations
but could not detect [C\,{\sc ii}] in individual pixels. However,
averaging the spectra above the $3\sigma$ contour in the FIFI-LS map
and S--W from M51b's center results in the spectrum shown in the lower
panel of Figure~\ref{fig:m51b_zoom}. The [C\,{\sc ii}] line is clearly
detected at the S--W with a peak at 580\,km\,s$^{-1}$, which is
consistent with the CO velocity in this region.  The total integrated
intensity of the observed spectrum is 2.4$\times
10^{-5}$\,erg\,s$^{-1}$\,cm$^{-2}$\,sr$^{-1}$, which is consistent
with the average intensity of the FIFI--LS map in this area of
2.0$\times 10^{-5}$\,erg\,s$^{-1}$\,cm$^{-2}$\,sr$^{-1}$.  We also
averaged the upGREAT spectra in the N--E part of the galaxy, but could
not detect any emission.

M51b is intrinsically very faint in [C\,{\sc ii}] considering the
observed TIR intensity and the inferred SFR.  Additionally, this
galaxy shows enhanced 24$\mu$m emission
(Figure~\ref{fig:cii_vs_tir_sfr}) and elevated dust temperature
\citep{MentuchCooper2012} compared to M51.  M51b is a barred
lenticular galaxy in a post--starburst phase, in which the stellar
population is dominated by old stars ($\gtrsim$10\,Gyr) and massive
star-formation is suppressed \citep{Kohno2002a,Alatalo2016}. The lack
of massive star formation in M51b is consistent with the faint
[C\,{\sc ii}] emission detected but it is inconsistent with the large
TIR emission observed in this galaxy and the high SFR inferred from
H$\alpha$ and 24$\mu$m dust continuum emission.

M51b shows Active Galactic Nuclei (AGN) activity, based on its
near--infrared line emission \citep{Goulding2009} and there is
evidence of AGN feedback producing arcs and shocks in X--ray and
H$\alpha$ emission \citep{Schlegel2016,Rampadarath2018}. X--rays can
heat the dust to higher temperatures than FUV emission, and can result
in bright infrared emission, resulting in enhanced values of the TIR
intensity and anomalously large inferred SFR \citep{Voit1991}. 
  Note, however, that the X-ray luminosity in M51b ($L_{\rm X}=5.4
  \times 10^{38}$\,erg\,s$^{-1}$) is low compared to what is typical
  for galaxies \citep{Ebrero2009}, and is a factor of 30 lower than
  that at the center of M51 \citep{Brightman2018}. Thus, the AGN in
  M51b can heat the dust only in its immediate vicinity.  Higher
  resolution images of the 24\,${\mu}$m and H$\alpha$ emission in M51b
  show a point source at its center that is unresolved in the
  6\arcsec\ (247\,pc) and 0.31\arcsec\ (12\,pc) resolution of these
  images, respectively. As the [C\,{\sc ii}] peak is about 10\arcsec\
  ($\sim$450\,pc) from the center of M51b, it is therefore possible
  that the bright fir-- and mid--infrared and [C\,{\sc ii}] emission
  arises from different locations within the 16\arcsec\ beam of our
  observations, with the hot dust associated to the AGN dominating the
  mid-- and far--infrared emission and a higher column density, colder
  dust temperature region contributing to the [C\,{\sc ii}]
  emission. This [C\,{\sc ii}]--emitting cloud is too far from the AGN
  to be influenced by its X--ray emission.
%
%
  The evolved stellar population in M51b can also heat the dust but
  does not contribute significantly to the FUV emission that heats the
  gas via photoelectric effect \citep{Bendo2012,Kapala2017}. Note,
  however, that the stellar surface density at the center of M51 is
  similar to that in M51b, but this region shows only a moderate
  deficit of [C\,{\sc ii}] with respect to the TIR emission and
  inferred SFR. Thus, an enhanced stellar density is likely not the
  only mechanism responsible for the low [C\,{\sc ii}]/SFR and
  [C\,{\sc ii}]/TIR ratios in M51b.

%
%
%
%

\subsection{[C\,{\sc ii}] emission at the center of M51}

\begin{figure}[t]
\plotone{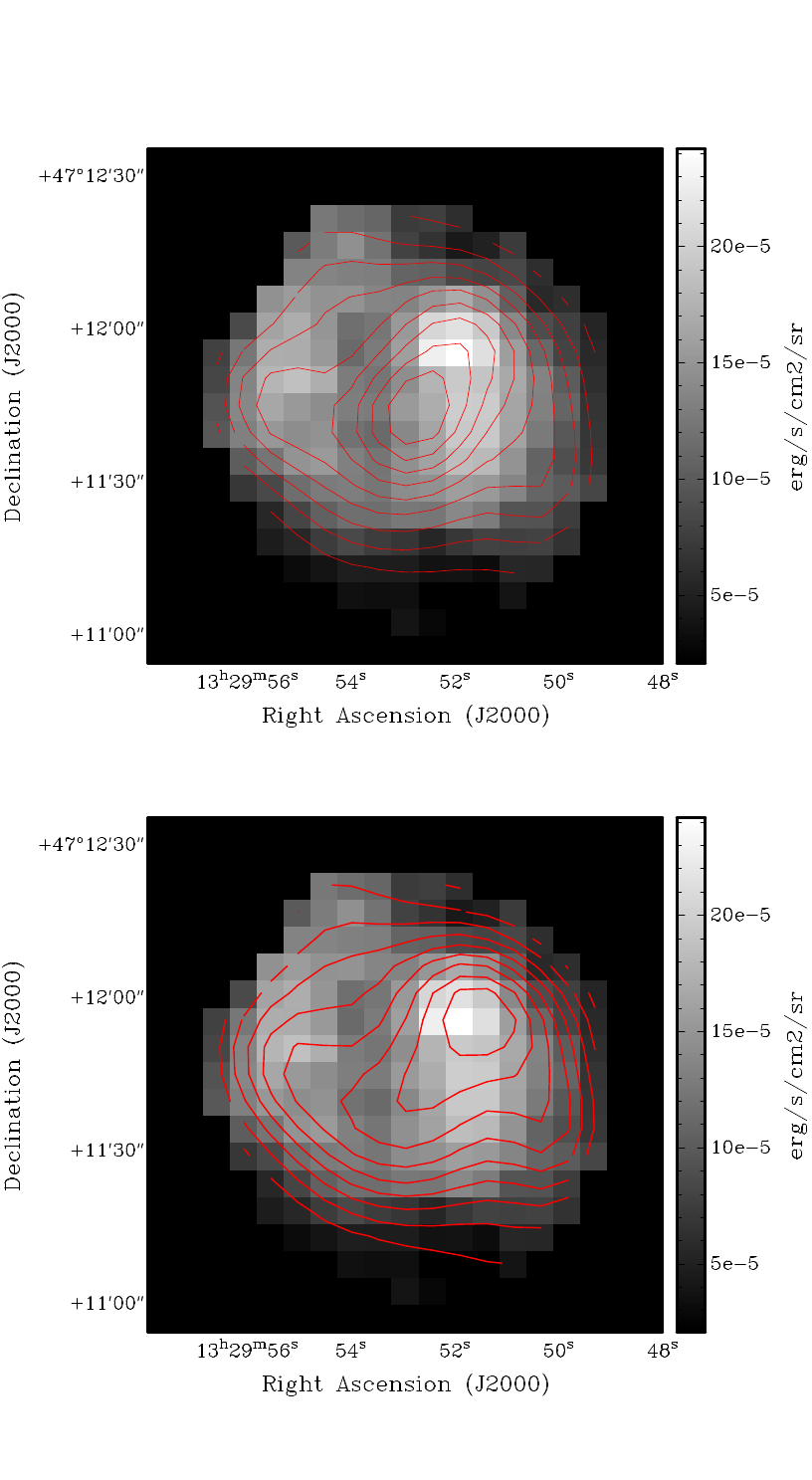}
\caption{({\it Upper panel}) Image of M51 center observed in [C\,{\sc
    ii}] with SOFIA/FIFI-LS overlaid with contours of the inferred SFR
  surface density. The contour lines  range from 10\% of the peak
  (0.22\,M$_{\odot}$\,yr$^{-1}$) to 90\% in steps of 10\%. ({\it Lower
    panel}) [C\,{\sc ii}] image of M51's center overlaid with contours
  of the TIR surface density. The contour lines correspond range from
  10\% of the peak (0.07 erg\,s$^{-1}$\,cm$^{-2}$\,sr$^{-1}$) to 90\%
  in steps of 10\%.\label{fig:center}}
\end{figure}

As shown in Figure~\ref{fig:cii_vs_tir_sfr}, there are a number of
pixels at the center of M51 that show a moderate [C\,{\sc ii}] deficit
for a given SFR surface density. However,  this effect is not as
pronounced when [C\,{\sc ii}] is compared with the TIR intensity. In
Figure~\ref{fig:center} we show the [C\,{\sc ii}] emission in M51's
central regions with contours of the TIR intensity and inferred SFR.
The SFR surface density image shows a peak of emission at its center
that is not present in the [C\,{\sc ii}] image (both the H$\alpha$ and
24\,$\mu$m show a point-source emission at this location ). This peak
corresponds to the location of the Compton--thick AGN activity present
in this region \citep{Stauffer1982,Fukazawa2001}. Therefore, this
region also represents an example of AGN--powered H$\alpha$ and
24\,$\mu$m emission, with fainter [C\,{\sc ii}] emission. The peak at
the AGN location can also be seen in the far--infrared images used to
estimate the TIR intensity but it is not as pronounced as for the
shorter wavelength data used to determine the SFR surface density.

\section{Conclusions}\label{sec:conclusions}

We presented a [C\,{\sc ii}] emission map over the entire M51 and M51b
galaxies observed with the SOFIA/FIFI--LS instrument. We compared the
[C\,{\sc ii}] emission with the total far--infrared intensity and star
formation rate surface density maps of M51 within a variety of
environments. We found that [C\,{\sc ii}] and the SFR surface density
are well correlated in the central, spiral arm, and inter-arm
regions. The correlation is in good agreement with that found for a
larger sample of nearby galaxies at kpc scales. We found that the SFR
and [C\,{\sc ii}] and TIR luminosities in M51 are dominated by the
extended emission in M51's disk.  The companion galaxy M51b, however,
shows a deficit of [C\,{\sc ii}] emission compared with the TIR
emission and inferred SFR surface density, with [C\,{\sc ii}] emission
detected only in the S--W part of this galaxy. We find that this
[C\,{\sc ii}] deficit is related to an enhanced dust temperature in
this galaxy.  We interpret the faint [C\,{\sc ii}] emission in M51b to
be a result of suppressed star formation in this galaxy, while the
bright mid-- and far--infrared emission, which drive the TIR and SFR
values, are powered by other mechanisms.  A similar but less
pronounced effect is seen at the location of the black hole in M51's
center. The observed [C\,{\sc ii}] deficit in M51b suggests that this
galaxy represents a valuable laboratory in which to study the origin
of the apparent [C\,{\sc ii}] deficit observed in ultra-luminous
infrared galaxies.

\acknowledgments Based on observations made with the NASA/DLR
Stratospheric Observatory for Infrared Astronomy (SOFIA). SOFIA is
jointly operated by the Universities Space Research Association,
Inc. (USRA), under NASA contract NAS2-97001, and the Deutsches SOFIA
Institut (DSI) under DLR contract 50 OK 0901 to the University of
Stuttgart.  We thank the staff of the SOFIA Science Center for their
help. We also thank an anonymous referee for a number of useful
comments that significantly improved the manuscript.  Part of the
research was carried out at the Jet Propulsion Laboratory, California
Institute of Technology, under a contract with the National
Aeronautics and Space Administration.  \copyright\ 2018. All rights
reserved. U.S. Government sponsorship acknowledged.

\vspace{5mm}
\facilities{SOFIA}

\bibliographystyle{aasjournal.bst}
\bibliography{papers}

\end{document}